\documentclass{article}

\usepackage{arxiv}

\usepackage[utf8]{inputenc} 
\usepackage[T1]{fontenc}    
\usepackage{hyperref}       
\usepackage{url}            
\usepackage{booktabs}       
\usepackage{amsfonts}       
\usepackage{nicefrac}       
\usepackage{microtype}      
\usepackage{lipsum}		
\usepackage{graphicx}
\usepackage{natbib}
\usepackage{doi}
\usepackage{siunitx}
\usepackage{amsmath,amssymb,amsfonts}
\usepackage{mathtools} 

\usepackage{amssymb}
\usepackage{pifont}
\usepackage{float}
\usepackage{caption}
\usepackage{subcaption}
\usepackage{comment}
\usepackage{mathtools}

\title{A Context-Switching/Dual-Context ROM Augmented RAM using Standard 8T SRAM}

\author{Md Abdullah-Al Kaiser\thanks{These authors contributed equally to this work.} \\
	University of Southern California\\
	Los Angeles, CA 90089 \\
	\texttt{mdabdull@usc.edu} \\
	\And
	Edwin Tieu$^*$ \\
	University of Southern California\\
	Los Angeles, CA 90089 \\
	\texttt{edwintie@usc.edu} \\
	\And
	Ajey P. Jacob \\
	Information Sciences Institute\\
	University of Southern California\\
	Los Angeles, CA 90292 \\
	\texttt{ajey@isi.edu} \\
 	\And
	Akhilesh R. Jaiswal \\
	University of Southern California\\
	Los Angeles, CA 90089 \\
	\texttt{akhilesh@usc.edu} \\
}

\renewcommand{\shorttitle}{CS/DC memory bit-cell}

\begin{document}
\maketitle

\begin{abstract}
The landscape of emerging applications has been continually widening, encompassing various data-intensive applications like artificial intelligence, machine learning, secure encryption, Internet-of-Things, etc. A \textit{sustainable} approach toward creating dedicated hardware platforms that can cater to multiple applications often requires the underlying hardware to context-switch or support more than one context simultaneously. This paper presents a \textit{context-switching} and \textit{dual-context} memory based on the standard 8T SRAM bit-cell. Specifically, we exploit the availability of multi-\si{V_T} transistors by selectively choosing the read-port transistors of the 8T SRAM cell to be either high-\si{V_T} or low-\si{V_T}. The 8T SRAM cell is thus augmented to store ROM data (represented as the \si{V_T} of the transistors constituting the read-port) while simultaneously storing RAM data. Further, we propose specific sensing methodologies such that the memory array can support RAM-only or ROM-only mode (context-switching (CS) mode) or RAM and ROM mode simultaneously (dual-context (DC) mode). Extensive Monte-Carlo simulations have verified the robustness of our proposed ROM-augmented CS/DC memory on the Globalfoundries 22nm-FDX technology node.
\end{abstract}



\maketitle

\section{Introduction}
According to Moore's law, the remarkable scaling of the Silicon transistor technology has driven steady improvements in power, performance, and area (PPA) metrics of state-of-the-art computing platforms \cite{shalf2020future}. The ever-improving hardware PPA metrics have been sustained through a series of innovations at device \cite{ye2019last}, circuit \cite{song202124}, and architectural level \cite{shin2018dnpu}. Historically, as the improvement in processor clock speed slowed down due to power concerns, parallel multi-core architectures emerged to cater to the ever-increasing compute demand of consumer applications \cite{leiserson2020there}. State-of-the-art hardware platforms feature multi-core architectures, forming the backbone of existing computing solutions.

Recently, however, the scope of consumer applications has vastly increased and is driven by data-intensive applications like big data, IoT, machine learning, artificial intelligence, secure encryption, etc. Coupled with the decreased pace of Moore's Law and the sky-rocketing compute demands of emerging applications, domain-specific architectures that can cater to the needs of a specific application of interest are being extensively explored by the research community \cite{jouppi2018domain}. Nevertheless, given the vast scope of emerging applications, hardware platforms often have to context switch between multiple applications. For example, dedicated domain-specific IoT devices could require support for both data analysis (machine learning) and data encryption (for secure wireless transfer) \cite{liu2021secdeep}. Thus, a sustainable pathway toward satisfying the computing need for a wide range of emerging applications requires custom hardware solutions that can seamlessly cater to multiple contexts while meeting the required power and performance metrics. Such dedicated multi-context hardware systems inevitably require an on-chip memory solution to rapidly context switch or cater to multi-context data. 

We present a Context-Switching (CS) and a Dual-Context (DC) on-chip memory solution based on standard 8T SRAM cells. Our proposal is based on augmenting the standard 8T SRAM cells \cite{verma2008256} with ROM-based memory. Each 8T bit-cell can simultaneously store independent RAM and ROM data, catering to multiple-context data stored within the same memory array. The ROM augmented part of the proposed scheme can store look-up tables for transcendental and polynomials functions for a wide range of applications or weights of a neural network \cite{fft, LUT, spare}, while the RAM part can serve as a scratch pad memory. Furthermore, the presented ROM-augmented RAM bit-cell can operate in context-switching (CS) mode, wherein the memory array can act as ROM or RAM array, or a dual-context (DC) mode, wherein the memory array can function as ROM and RAM array, simultaneously. Our proposed bit-cell can perform better for \text{ROM-intensive} workloads that require frequent ROM access and computation that depends heavily on ROMs. Prior works on ROM-embedded SRAM can cater to ROM or RAM data, one at a time and/or use increased wordline/bit-line capacitance or multiple supply rails that can degrade speed and energy-efficiency \cite{area_ROM, RECache, multi_supply, multi_bitline}.  In contrast, our proposal can cater to both CS and DC modes, wherein RAM and ROM data can be accessed simultaneously by exploiting advanced foundry nodes' multi-\si{V_T} nature. 

The key highlights of the paper are as follows:
\begin{enumerate}
    \item We present a novel approach to augment standard 8T SRAM cells with ROM, such that a single bit-cell can simultaneously store RAM and ROM data.
    
    \item We propose two different operating modes for the presented ROM augmented RAM - a) Context-Switching (CS) Mode, wherein the 8T SRAM array can operate either as RAM \textit{or} ROM, b) Dual-Context (DC) Mode, wherein the memory array can simultaneously operate both as a RAM \textit{and} ROM. 
    
    \item For the DC mode of operation, we propose using a single sense amplifier with a dual-thresholding sensing scheme for reading both RAM and ROM data. In addition, in DC mode, our proposed method can read both RAM and ROM data within two cycles without destructing or storing data into a temporary buffer, thus saving latency and power overhead. 
\end{enumerate}

\section{Proposed Context-Switching / Dual-Context 8T SRAM Bit-Cell}

\subsection{Proposed Bit-cell}

\begin{figure}[!t]
\centering
\subfloat[ROM data = `0']{\includegraphics[width=0.5\linewidth]{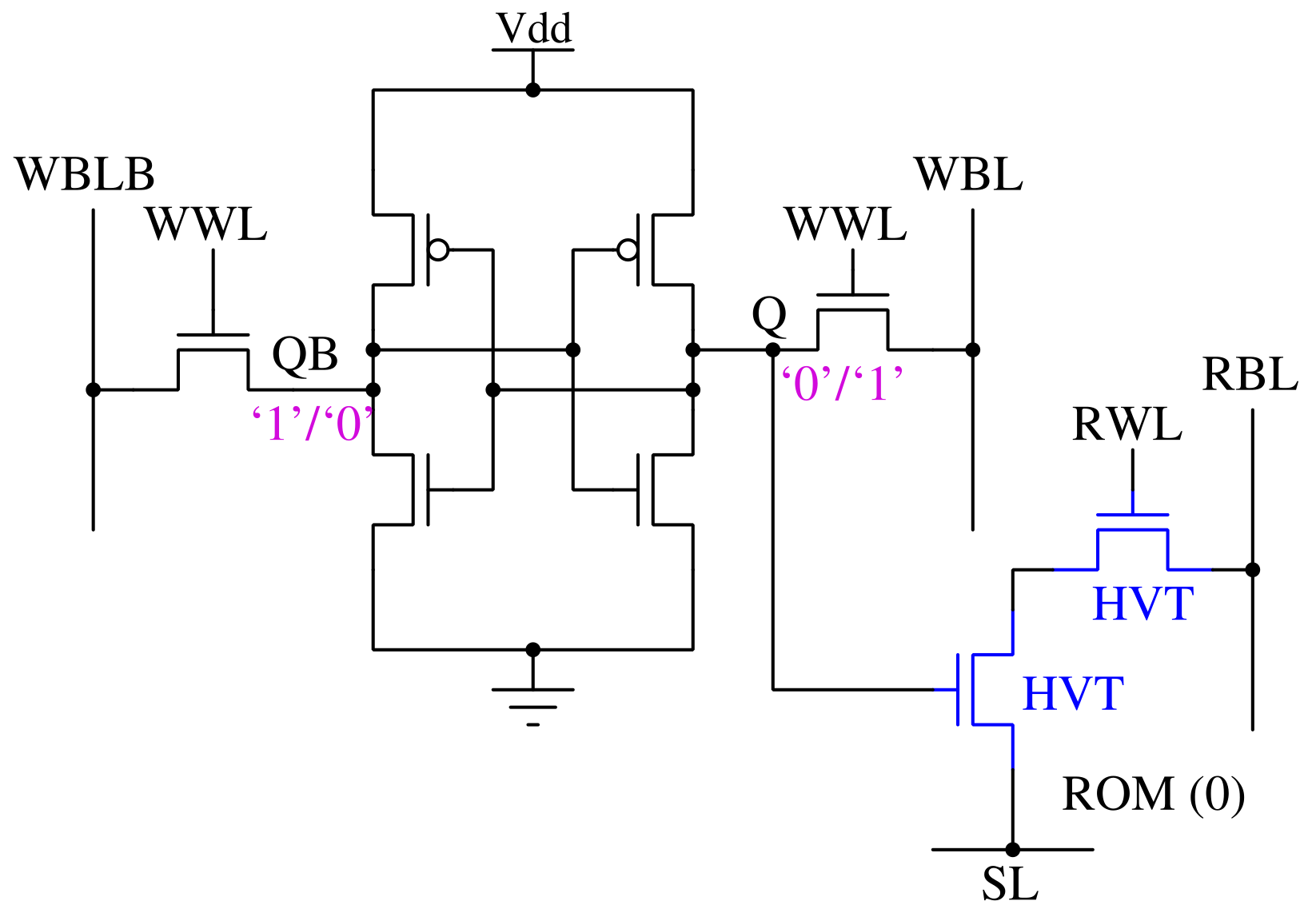}} 
\subfloat[ROM data = `1']{\includegraphics[width=0.5\linewidth]{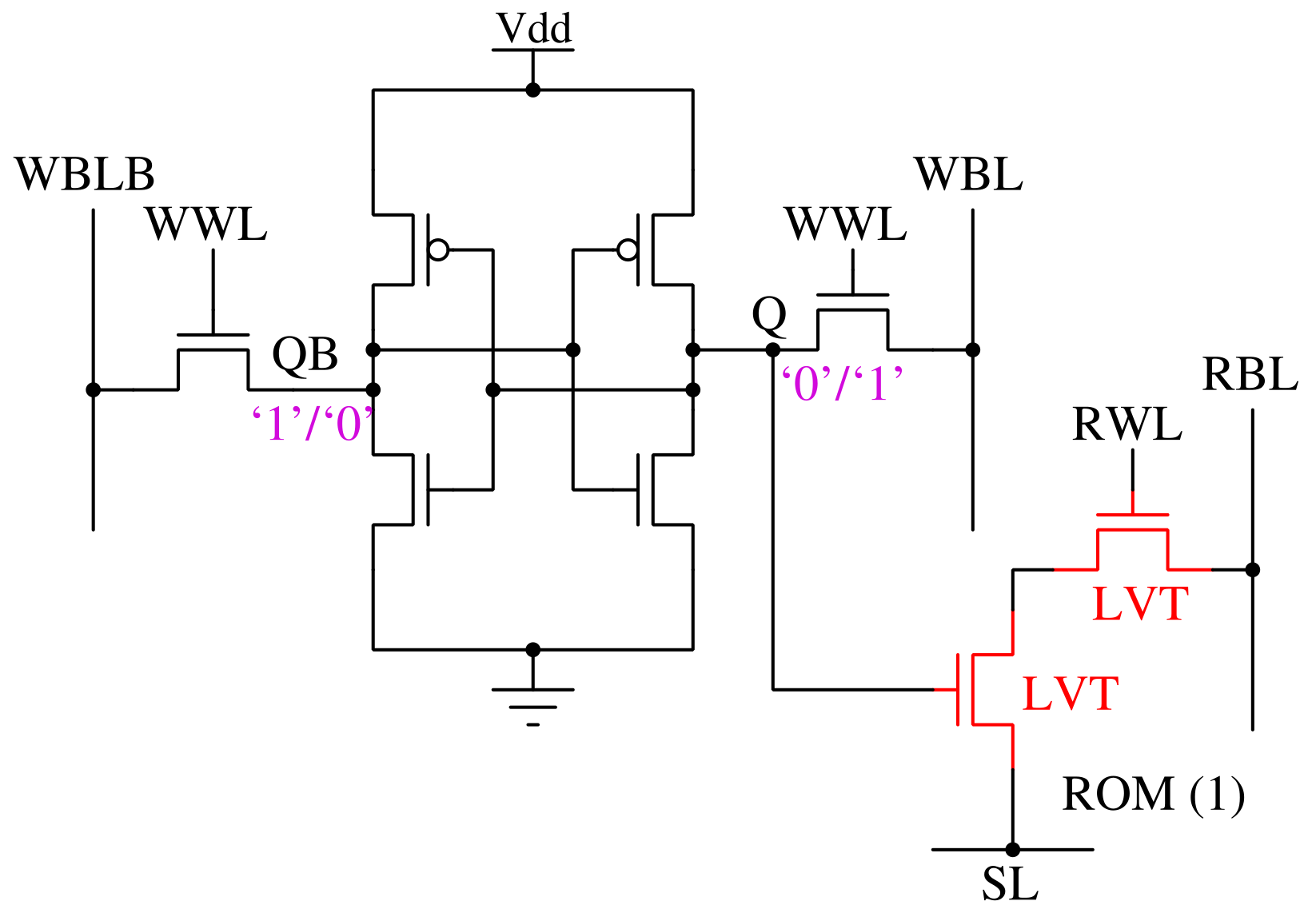}}
\caption{Proposed Context-Switching/ Dual-Context memory using standard 8T SRAM bit-cell. High-\si{V_T} and low-\si{V_T} read transistors represent the ROM data (a) `0' and (b) `1', respectively.}
\label{bitcell_schematic}
\end{figure}

Figure \ref{bitcell_schematic} shows the proposed CS/DC memory bit-cell. It consists of the standard 8T SRAM bit-cell with a decoupled read-port. Write operation is achieved by utilizing the write-port (WWL, WBL, and WBLB) similar to the standard 6T SRAM. However, the read operation is performed through the decoupled read-port (RWL, RBL) and shared source line (SL). To embed ROM data inside the standard 8T SRAM cell, we propose to exploit the availability of multi-\si{V_T} transistors in commercial foundry process design kits (PDKs). Specifically, the read-port of the standard 8T SRAM cell can be constructed by either using high-\si{V_T} or low-\si{V_T} transistors. The high-\si{V_T} and low-\si{V_T} transistors represent the ROM data bit of `0' and `1', respectively.  Figure \ref{bitcell_schematic}(a), and \ref{bitcell_schematic}(b) illustrate the CS/DC bit-cells storing the ROM data bit of `0' and `1', respectively. As shown in the figure, when the read-port transistors are implemented using high-\si{V_T} (low-\si{V_T}) transistors, the 8T bit-cell can be considered to be storing a ROM data `0' (ROM data `1'), in addition to the usual SRAM data stored on nodes Q and QB. Thus, the presented 8T bit-cell, wherein the read-port transistors are selectively chosen to be either high-\si{V_T} or low-\si{V_T}, simultaneously stores one-bit RAM data and one-bit ROM data \textit{within the same memory footprint}.

\subsection{RAM and ROM Read Analysis}

\begin{figure*}[!b]
\centering
\subfloat[\si{V_{SL}} = -0.10 V]{\includegraphics[width=0.3\linewidth]{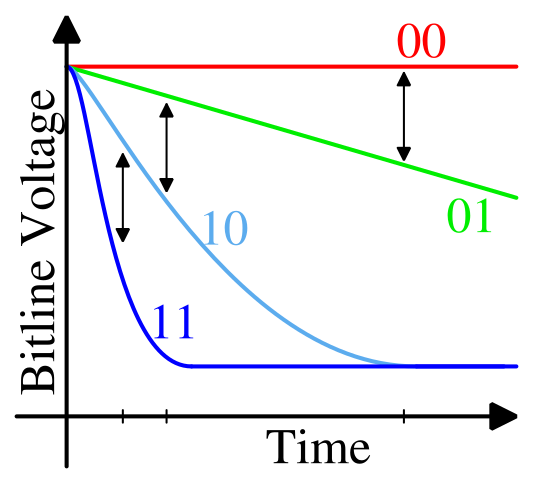}}
\subfloat[\si{V_{SL}} = -0.10 V]{\includegraphics[width=0.5\linewidth]{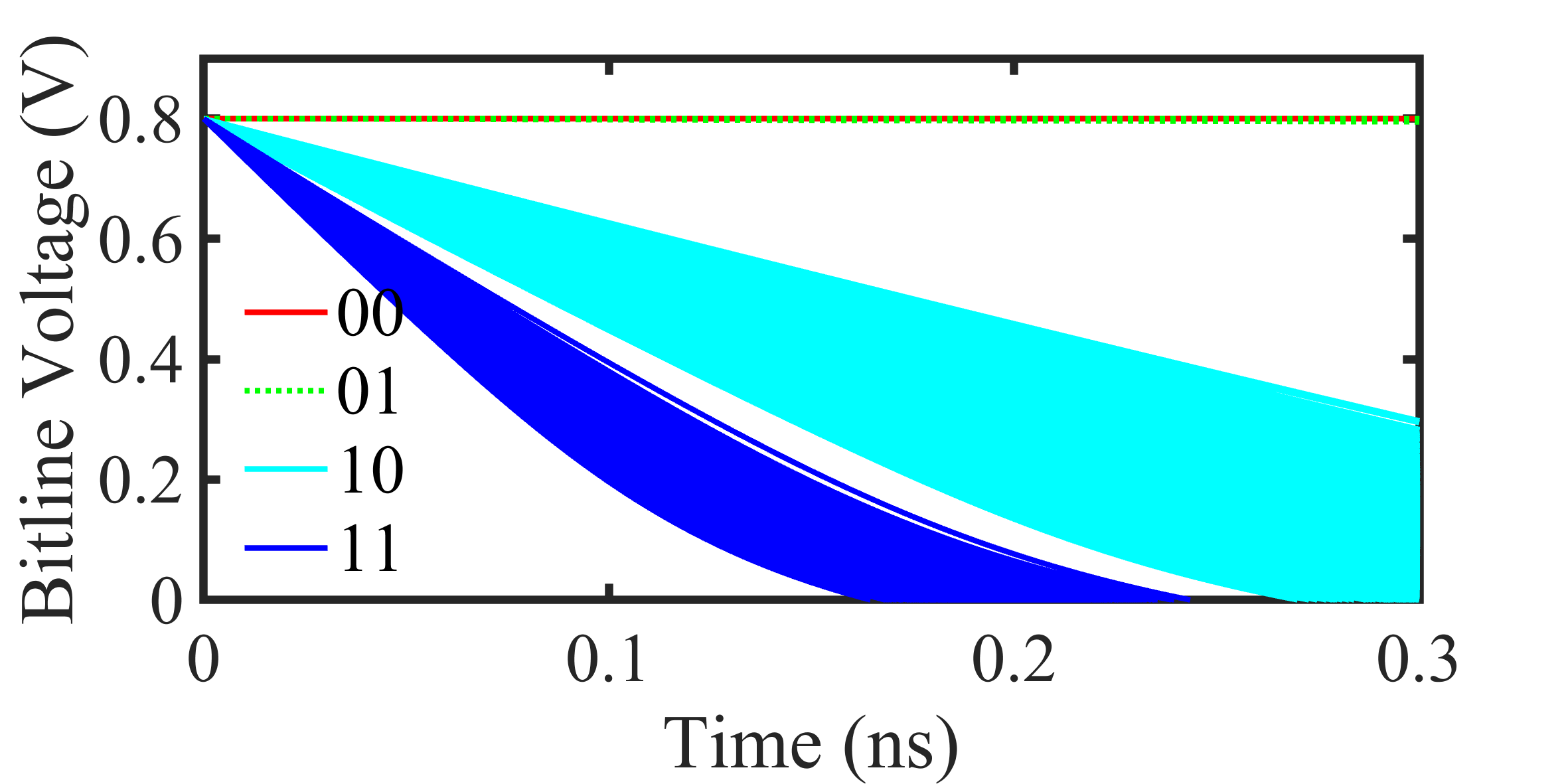}} \\
\subfloat[\si{V_{SL}} = +0.20 V]{\includegraphics[width=0.5\linewidth]{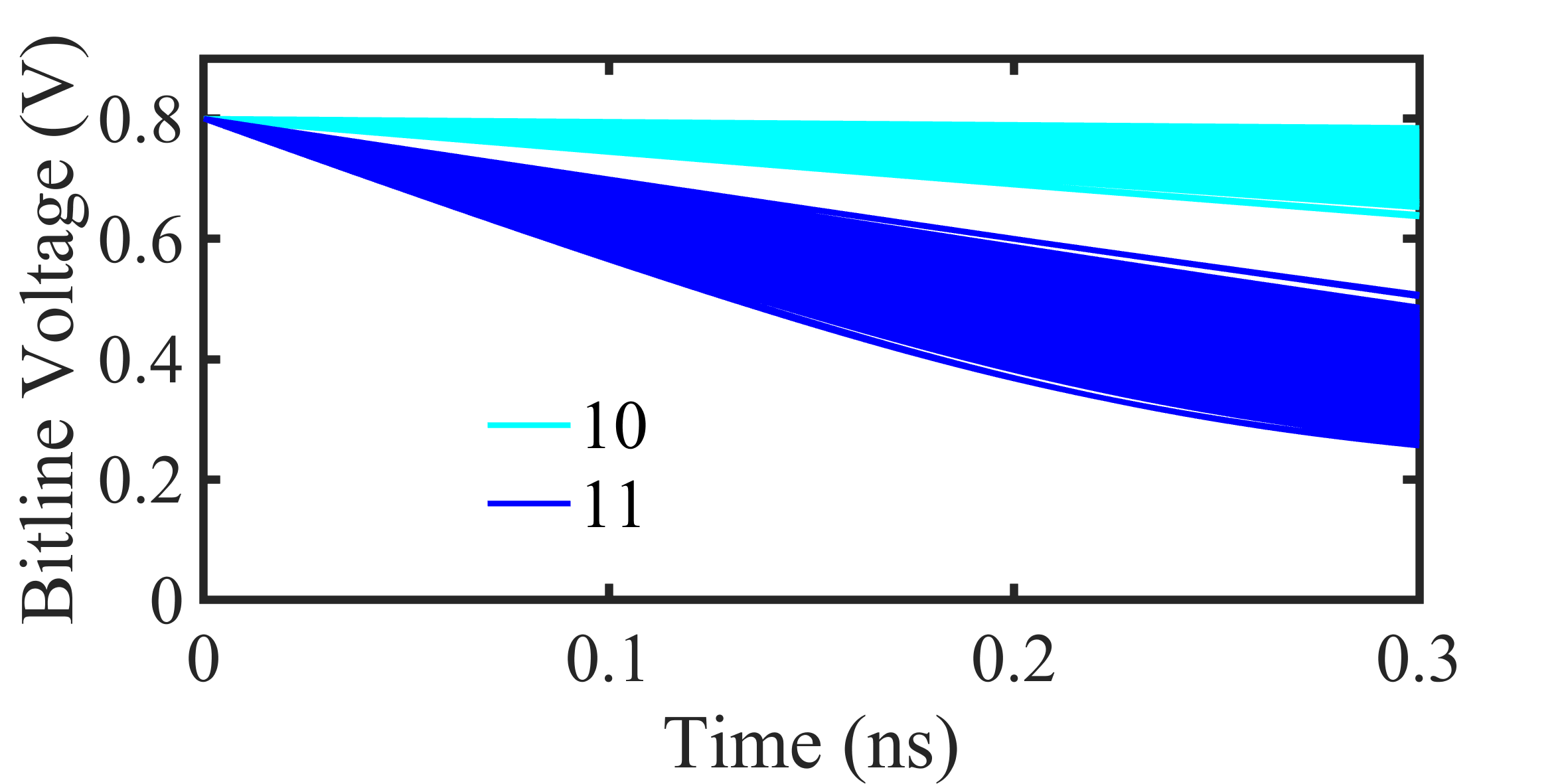}}
\subfloat[\si{V_{SL}} = -0.45 V]{\includegraphics[width=0.5\linewidth]{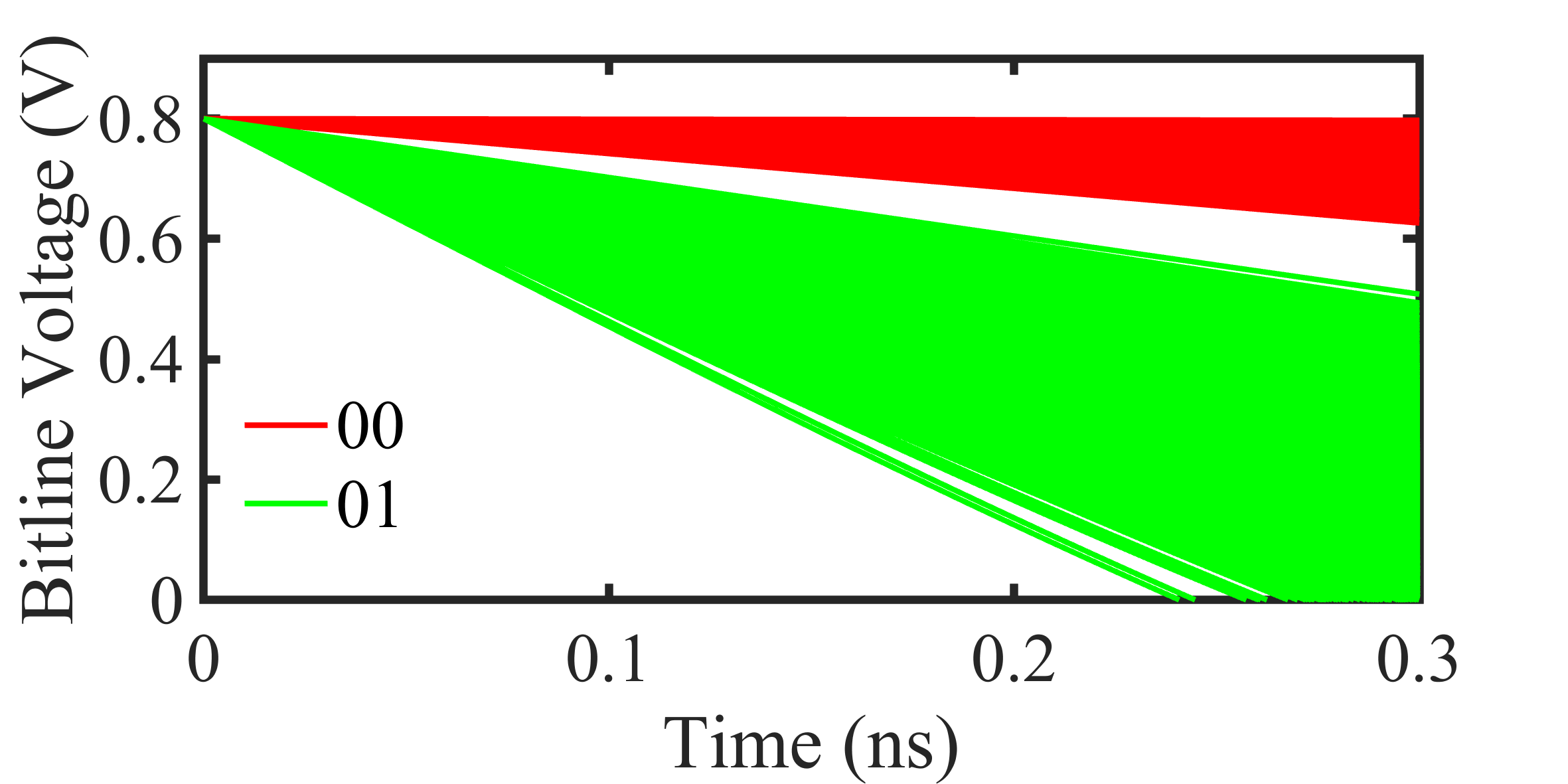}}
\caption{(a) Conceptual bit-line discharge voltage versus time for different data bit combinations in the proposed CS/DC bit-cell. MSB and LSB represent the SRAM and ROM stored data bit, respectively, (b), (c), and (d) Monte-Carlo simulations of the bit-line discharge voltage versus time for different source line (SL) voltages at TT corner for 5000 samples per each case.}
\label{mc_voltage_discharge}
\end{figure*}

To read the RAM and the ROM data stored in the 8T bit-cell, the sensing circuit in the periphery should be able to differentiate between various BL discharge rates controlled by the node Q (stored SRAM data) and the \si{V_T} flavor of the transistor constituting the read-port (stored ROM data). Figure \ref{mc_voltage_discharge}(a) illustrates a typical bit-line discharge voltage versus time for different SRAM and ROM data bit combinations. The proposed CS/DC bit-cell can store 2-bit, 1-bit re-writable data in the SRAM and 1-bit fixed data in the ROM. For the rest of the paper, we follow the convention that the MSB and LSB represent the SRAM and ROM data bits stored in the same 8T bit-cell, respectively. 

Consider the source line (SL) is connected to a small negative voltage within the reliability limit of the transistors. For case-00 (i.e., Q = 0 and the read-port transistors are high-\si{V_T}), the pre-charged read bit-line will remain close to \si{V_{DD}}. The negative \si{V_{SL}} and the SRAM stored data bit (Q = 0) cannot turn ON the high-\si{V_T} transistor along the read path. However, for case-01 (i.e., Q = 0 and the read-port transistors are low-\si{V_T}), the bit-line would experience slow discharge. Note, though the SRAM data bit is `0', the negative \si{V_{SL}} marginally activates the low-\si{V_T} read transistor. On the other hand, for case-10 (i.e., Q = 1 and the read-port transistors are high-\si{V_T}) and case-11 (i.e., Q = 1 and the read-port transistors are low-\si{V_T}), the bit-line voltage discharges faster compared to the Case-01 due to higher gate-voltage-overdrive of the lower read-port transistor. Further, the high-\si{V_T} read transistors (Case-10) will have slower discharge compared to the low-\si{V_T} (Case-11) transistors. Thus, various RAM and ROM data stored in the same 8T bit-cell lead to different rates of BL discharge. The sensing circuit can read the ROM and RAM data bits from the same bit-cell. 

Figure \ref{mc_voltage_discharge}(b), (c), and (d) show the bit-line discharge voltage versus time considering the local variation at TT (Typical) corner for 5000 simulations for different  source line votlages per each data set. Figure \ref{mc_voltage_discharge}(b) exhibits that there is no robust sense margin between case-00 and case-01, similarly, between case-10 and case-11 due to the local mismatch variation when the SL is connected to a small negative voltage (i.e., -0.1 V). Hence, applying the small negative \si{V_{SL}} can only differentiate between the SRAM data; however, it cannot sense the ROM data robustly. Figure \ref{mc_voltage_discharge}(c) shows a sufficient sense margin between case-10 and case-11 when the \si{V_{SL}} is positive (i.e., 0.2 V). Due to the application of the positive \si{V_{SL}} with the stored SRAM data `1', the high-\si{V_T} read-port transistor can marginally start conducting. In contrast, the low-\si{V_T} transistor can entirely turn on due to its lower threshold voltage requirement. As a result, the sensing circuit can easily differentiate between case-10 and case-11. Moreover, Figure \ref{mc_voltage_discharge}(d) exhibits that the necessary sense margin can be achieved between case-00 and case-01 by applying a larger negative voltage (\si{V_{GS}} < \si{V_{DD}} as \si{V_G} of the read transistor is 0 V for case-00 and case-01) at the source line (SL). The applied \si{V_{GS}} for the lower read-port transistors exceeds the threshold voltage of the low-\si{V_T} transistor; however, it is not enough to completely turn on the high-\si{V_T} transistor. As a result, ROM data can be easily differentiated when SRAM stored data is `0' by applying the appropriate negative \si{V_{SL}}.

\section{Mode of Operation}

The proposed scheme can operate in both context-switching and dual-context modes robustly. The memory array supports RAM or ROM read operation in the context-switching mode. Conversely, SRAM and ROM data can be read simultaneously in the dual-context mode. Simultaneous SRAM and ROM data access is achieved in two phases by using proper \si{V_{SL}} to detect 2-bit data. Detailed discussions about the sensing operation with timing waveforms for both modes are described below.

\subsection{Context-Switching Mode}

\begin{figure}[!b]
\centering
\subfloat[ROM-only]{\includegraphics[width=0.32\linewidth]{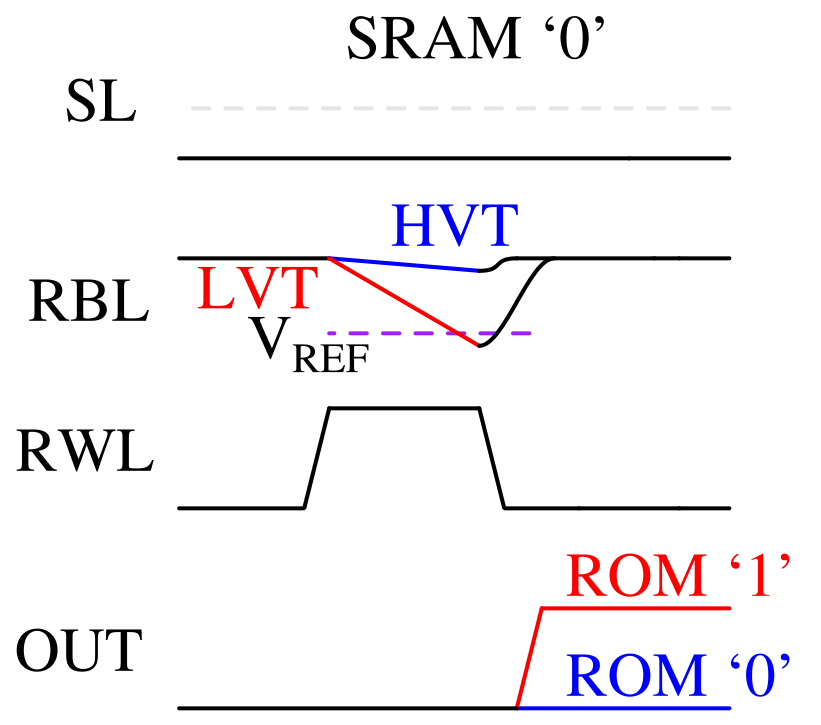}}
\subfloat[RAM-only]{\includegraphics[width=0.32\linewidth]{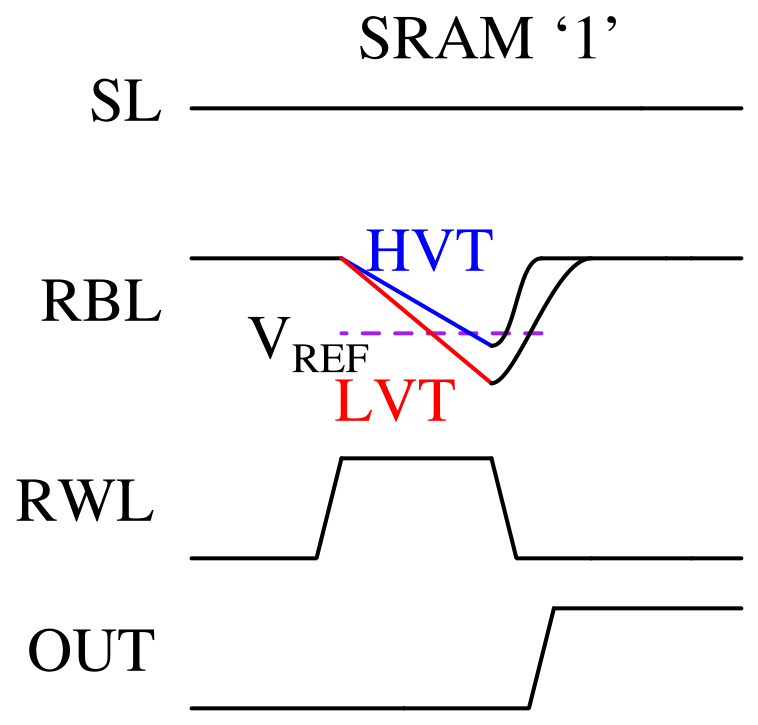}}
\subfloat[RAM-only]{\includegraphics[width=0.32\linewidth]{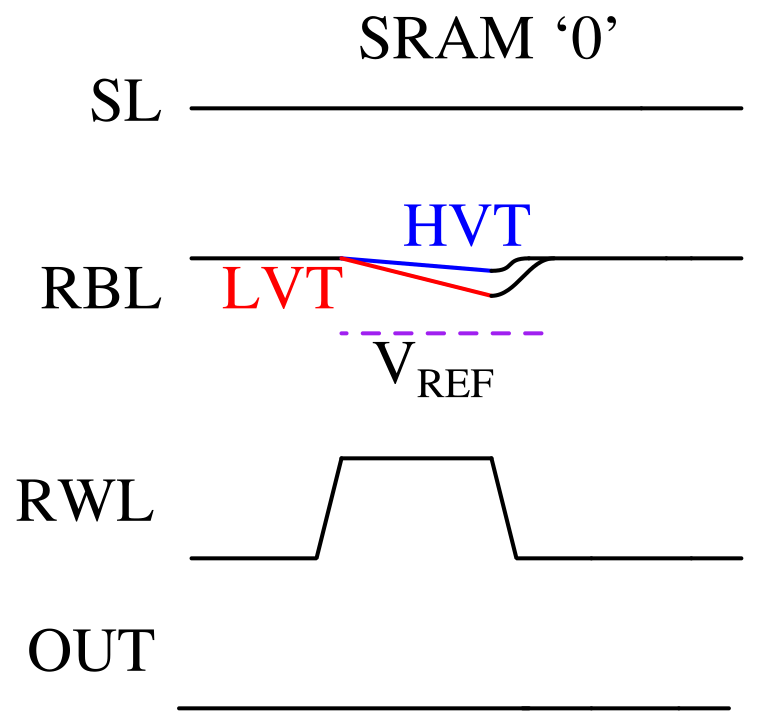}}
\caption{Timing waveform of the (a) ROM-only mode sensing when SRAM data is `0', (b) and (c) RAM-only mode sensing when SRAM data is `1' and `0', respectively for both low-\si{V_T} and high-\si{V_T} read-port transistors.}
\label{CS_wv}
\end{figure}

\subsubsection{ROM-only Mode}
During the ROM-only mode, the array acts as a ROM array and stores only the ROM data. All the SRAM bit-cells are initialized such that Q = `0' for all SRAM cells, and a negative voltage is applied at the source line (SL). When the RWL is activated, the bit-line voltage discharges faster for the low-\si{V_T} read transistors than the high-\si{V_T} read transistors. The negative \si{V_{SL}} is chosen in such a way that the effective \si{V_{GS}} of the lower read-port transistor is sufficient to turn on the low-\si{V_T} read-port transistor, whereas the high-\si{V_T} transistor is marginally on. A standard current-based sense amplifier with the appropriate reference voltage can sense the stored ROM data. When the ROM data is `0' (high-\si{V_T}), the sense amplifier outputs 0, and when the  ROM data is `1' (low-\si{V_T}), the output becomes 1. Figure \ref{CS_wv}(a) shows the timing waveform for the ROM-only mode operation.  

\subsubsection{RAM-only Mode}
ROM data can be both `1' and `0' in the RAM-only read operation. Hence, SRAM data needs to be detected irrespective of the low-\si{V_T} or high-\si{V_T} read-port transistors. The source line (SL) can be grounded like standard SRAM sensing in this mode. Note the source line (SL) can also be pulled down to a negative voltage (within the voltage reliability limit of the transistors) for faster sensing. The bit-line discharges below a specific reference voltage when the SRAM data is `1' for low-\si{V_T} or high-\si{V_T} read transistors. In contrast, the bit-line remains close to \si{V_{DD}} when the SRAM data is `0'. Hence,  a sense amplifier can differentiate between the SRAM data `1' and `0' with an appropriate reference voltage. Figure \ref{CS_wv}(b) and \ref{CS_wv}(c) show the timing waveform for the RAM-only mode sensing when SRAM is storing data `1' and `0', respectively. 

\begin{figure}[!b]
  \centering
  \includegraphics[width=0.9\linewidth]{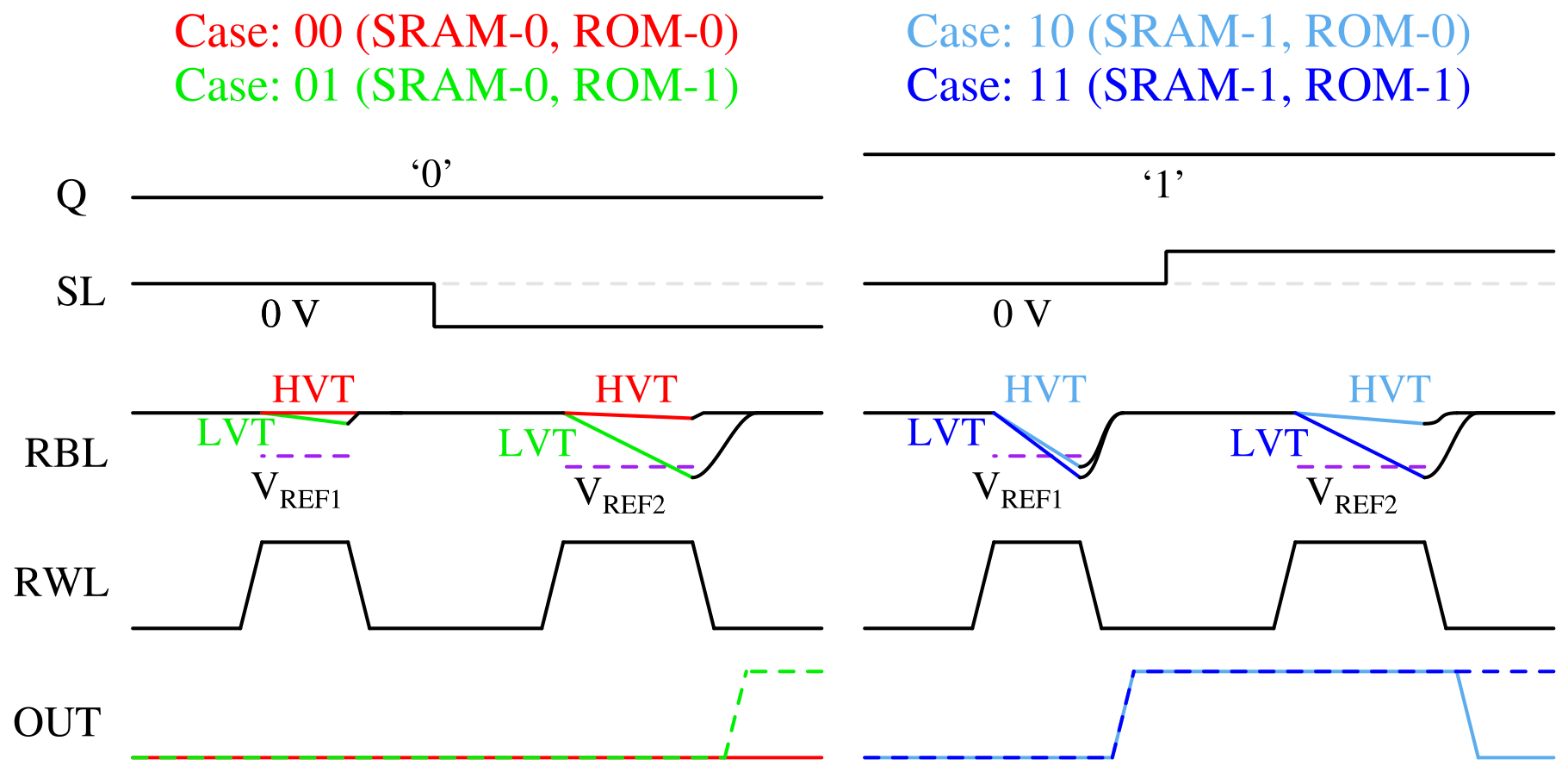}
  \caption{Timing waveform of the Dual-Context mode sensing.}
  \label{dual_mode_wv}
\end{figure}

\subsection{Dual-Context Mode}

The dual-context mode performs the read operation utilizing two phases and a single sense amplifier to detect 2-bit of data simultaneously without destroying the SRAM data. In the first phase, the source line (SL) node will be connected to the ground, and the standard SRAM sensing operation will be performed. In the next phase, the control circuit will connect the shared source line (SL) to a positive or negative voltage based on the SRAM data. Figure \ref{dual_mode_wv} illustrates the timing waveform for all 2-bit data combinations. For case-00 and case-01, the sense amplifier detects the SRAM data in the first phase; as the SRAM data is `0', the control circuit will connect the source line (SL) to a negative voltage for the next phase. Figure 3(d) shows that ROM data can be differentiated by applying proper reference voltage on the RBL when the SL is pulled to a negative voltage. The effective \si{V_{GS}} applied at the lower read-port transistor is enough to turn on the low-\si{V_T} transistor; however, not sufficient for the high-\si{V_T} transistor. As a result, the low-\si{V_T} read transistor will allow the bit-line to discharge below the reference voltage, generating high logic output (case-01). In contrast, the bit-line voltage remains close to \si{V_{DD}} when the read-port is constituted of a high-\si{V_T} transistor (case-00). 
Similarly, for case-10 and case-11, the control circuit will connect the source line (SL) with a positive voltage as the SRAM is now storing data `1'. Note, as shown in Figure 3(c), for case-10, the bit-line voltage cannot discharge below the reference voltage due to the high-\si{V_T} read-port transistors; in contrast, the bit-line voltage goes below the reference for the case-11 due to the low-\si{V_T} read-port transistors. Thus, a two-phase operation wherein, in phase-I, the sensing circuit determines the data stored in the SRAM cell, and in phase II, the sensing circuit can use appropriate source line (SL) voltage to determine the ROM data, achieving a dual-context operation.

\section{Evaluation and Process Variation Analysis}

The proposed CS/DC memory bit-cell has been implemented using 22nm Globalfoundries FDSOI technology. To verify the functionality in CS and DC mode, Monte-Carlo simulations have been run for 5000 samples per each case, considering the local variation at the TT corner only using HSPICE. The global PVT variation effects can be calibrated by adjusting the read word line (RWL) delay, reference voltages of the sense amplifier, and source line (SL) voltage, hence, ignored for the verification.  

\begin{figure}[!t]
\centering
\subfloat[ROM data = `0']{\includegraphics[width=0.5\linewidth]{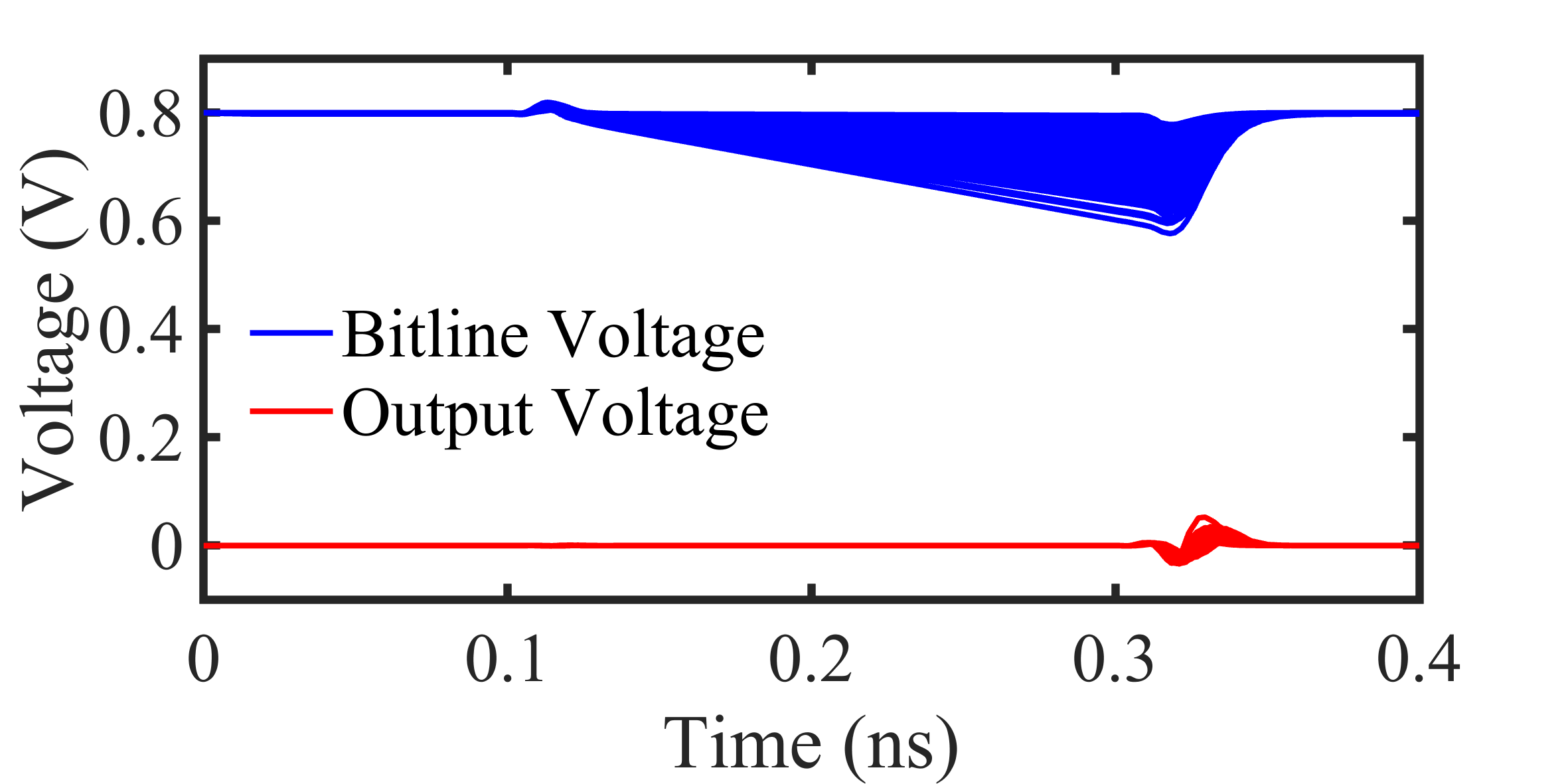}}
\subfloat[ROM data = `1']{\includegraphics[width=0.5\linewidth]{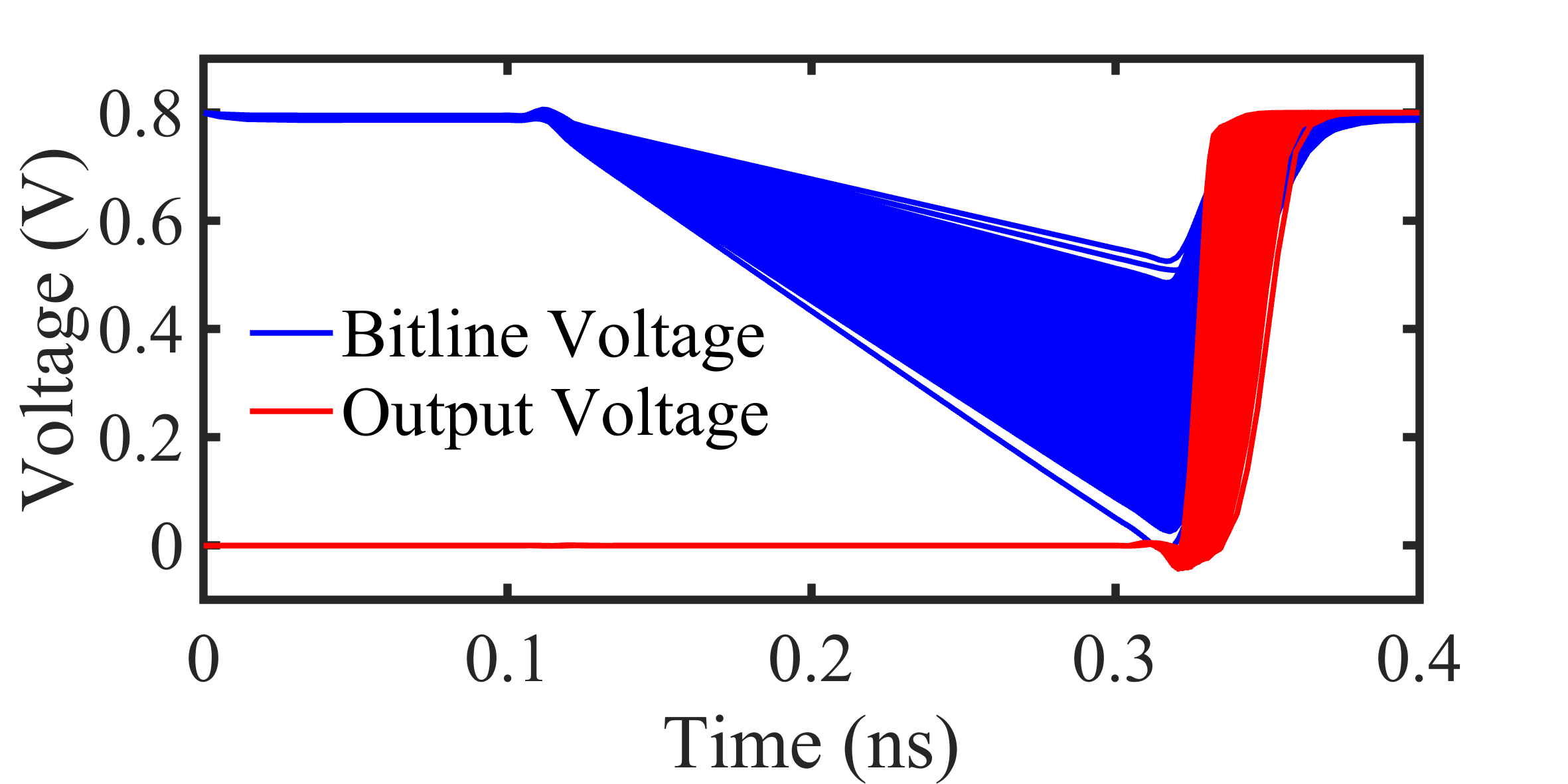}}
\caption{Monte-Carlo simulation results at TT corner for ROM-only mode.}
\label{mc_rom_only_mode}
\end{figure}

\begin{figure}[!b]
\centering
\subfloat[RAM data = `0']{\includegraphics[width=0.5\linewidth]{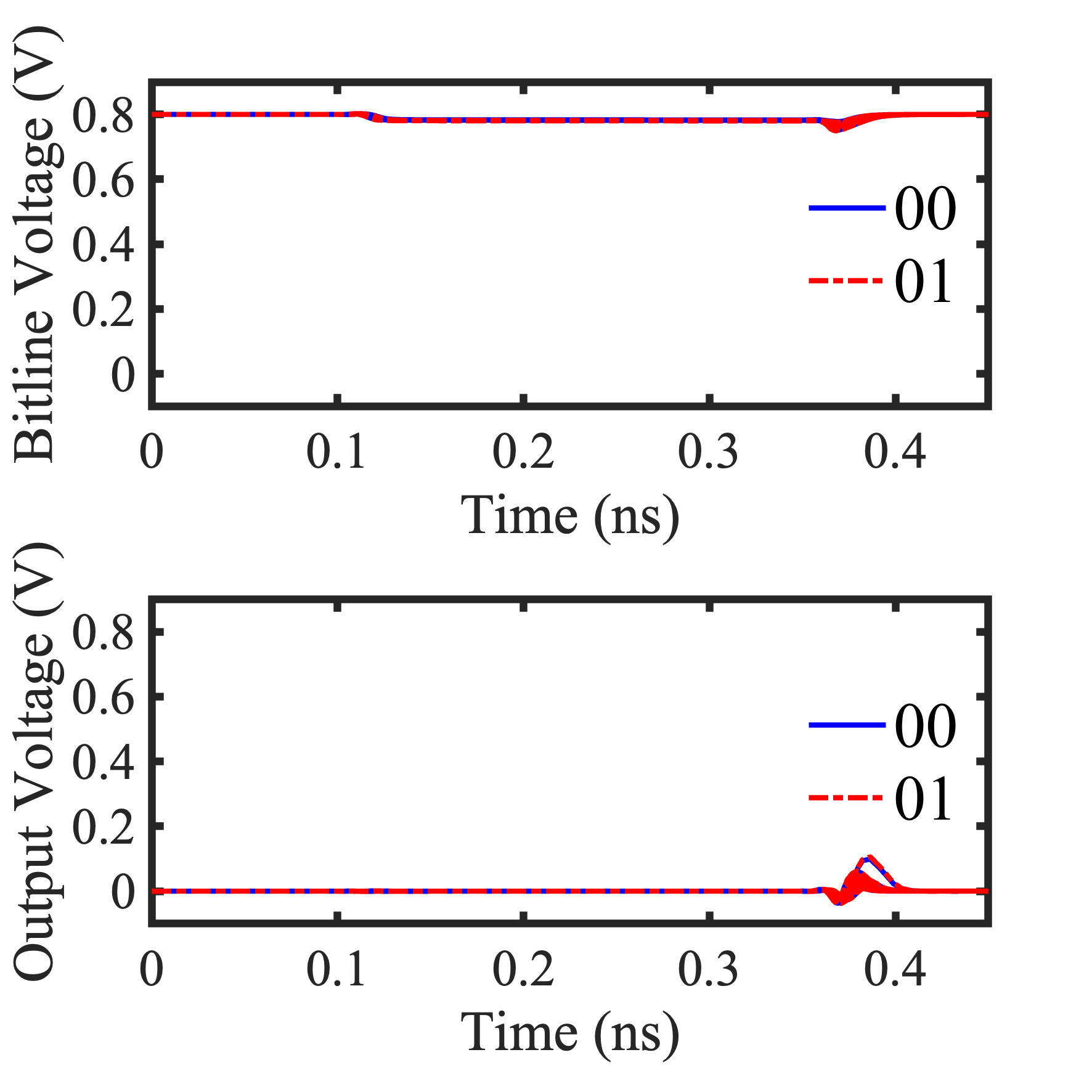}}
\subfloat[RAM data = `1']{\includegraphics[width=0.5\linewidth]{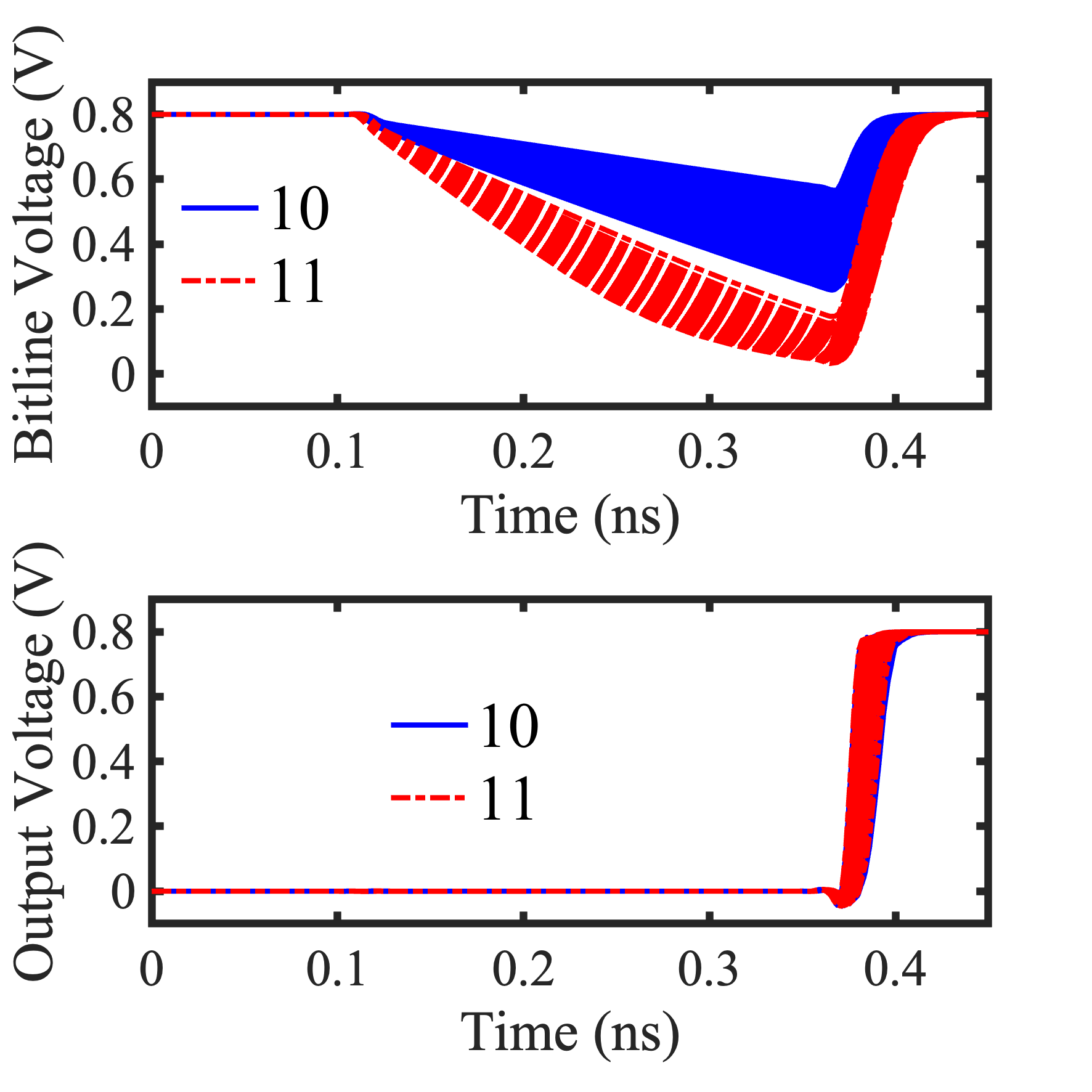}}
\caption{Monte-Carlo simulation results at TT corner for the reliability-friendly RAM-only mode when the source line (SL) is connected to the ground.}
\label{mc_ram_only_mode}
\end{figure}

\begin{figure}[!t]
\centering
\subfloat[RAM data = `0']{\includegraphics[width=0.5\linewidth]{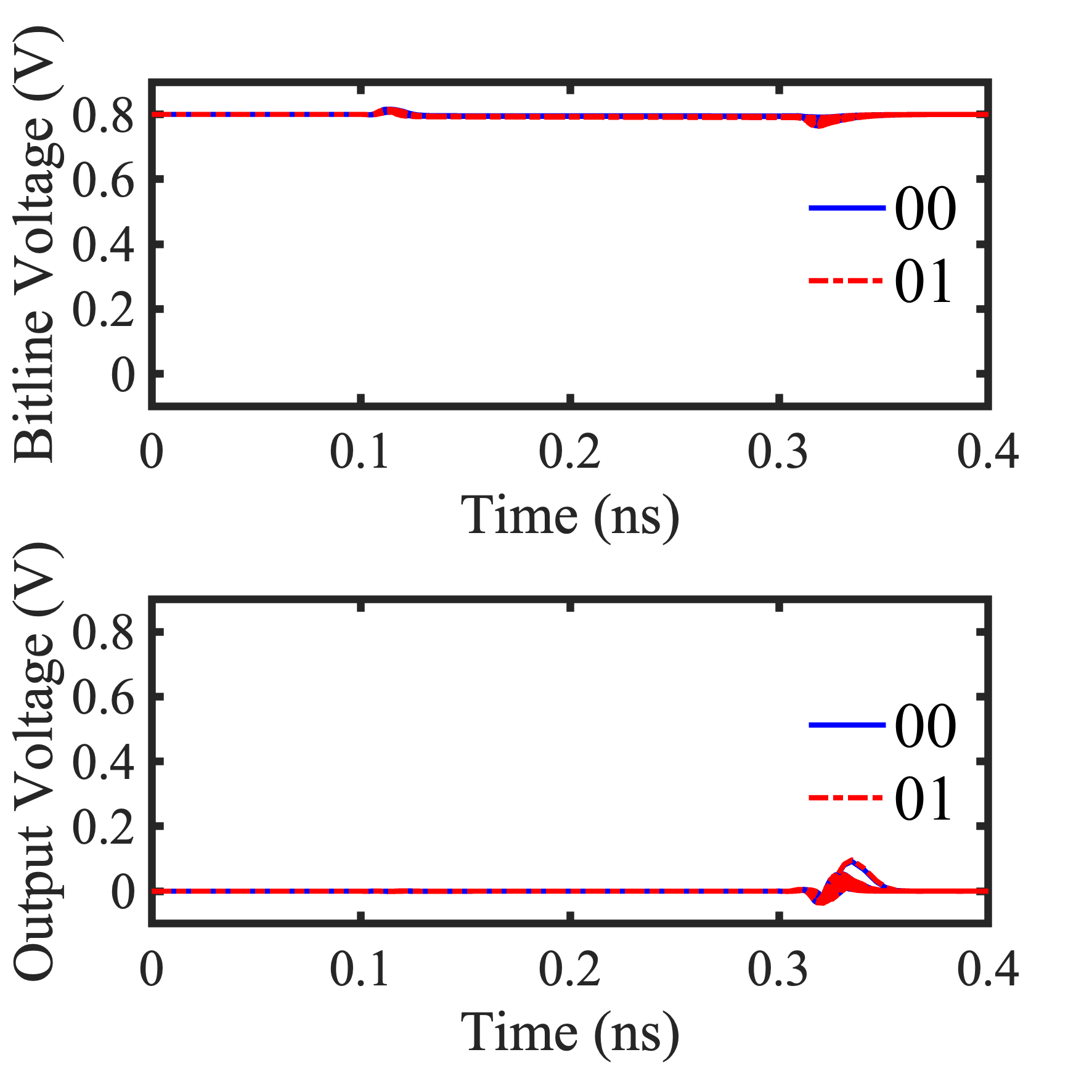}}
\subfloat[RAM data = `1']{\includegraphics[width=0.5\linewidth]{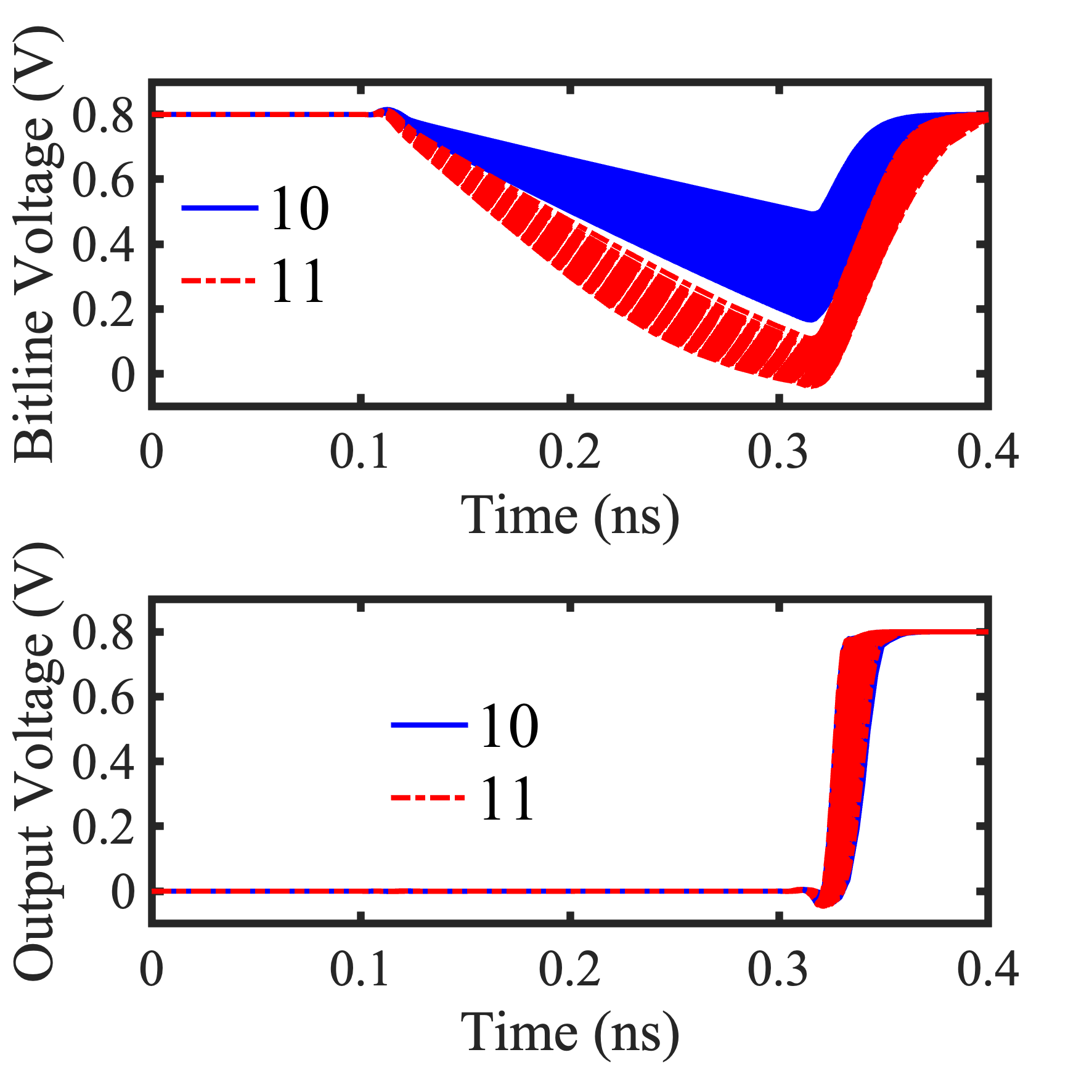}}
\caption{Monte-Carlo simulation results at TT corner for the delay-friendly RAM-only mode when the source line (SL) is connected to a negative voltage.}
\label{mc_ram_only_mode_negsl}
\end{figure}

\subsection{Context-Switching Mode}

\subsubsection{ROM-only Mode}
Figure \ref{mc_rom_only_mode} shows the bit-line voltage and the output of the sense amplifier for 5000 samples for both ROM data of `0' and `1' (high-\si{V_T} and low-\si{V_T} read-port transistors). A noticeable sense margin can be observed from the figure between the ROM data `0' and `1', and the sense amplifier can robustly read the ROM data.

\subsubsection{Reliability-friendly RAM-only Mode}
Figure \ref{mc_ram_only_mode} shows Monte-Carlo simulation results in the reliability-friendly RAM-only mode. In this mode, the source line (SL) is connected to the ground; as a result, the \si{V_{GS}} of the lower transistor of the read-port never exceeds the \si{V_{DD}}. The figure shows that the bit-line voltage discharges faster with a different rate and goes below the reference voltage when the SRAM is storing `1' irrespective of the low-\si{V_T} or high-\si{V_T} read-port transistors. In contrast, the bit-line voltage remains close to \si{V_{DD}} when SRAM stores `0'. 

\subsubsection{Delay-friendly RAM-only Mode}
Figure \ref{mc_ram_only_mode_negsl} shows the Monte-Carlo simulation results in the delay-friendly RAM-only mode when the source line (SL) is connected to a negative voltage. As a result, when the SRAM is storing `1', the \si{V_{GS}} across the lower read-port transistor exceeds \si{V_{DD}}; hence, the bit-line discharges at a  faster rate than the reliability-friendly mode due to the gate overdrive. This delay-friendly mode can compensate for the extra delay associated with the high-\si{V_T} read-port transistors of our CS/DC bit-cell. It can be illustrated from Figure \ref{mc_ram_only_mode_negsl} that the output becomes high (low) when SRAM is storing `1' (`0') regardless of the stored ROM data.

\begin{figure*}[!t]
\centering
\subfloat[case-00]{\includegraphics[width=0.5\linewidth]{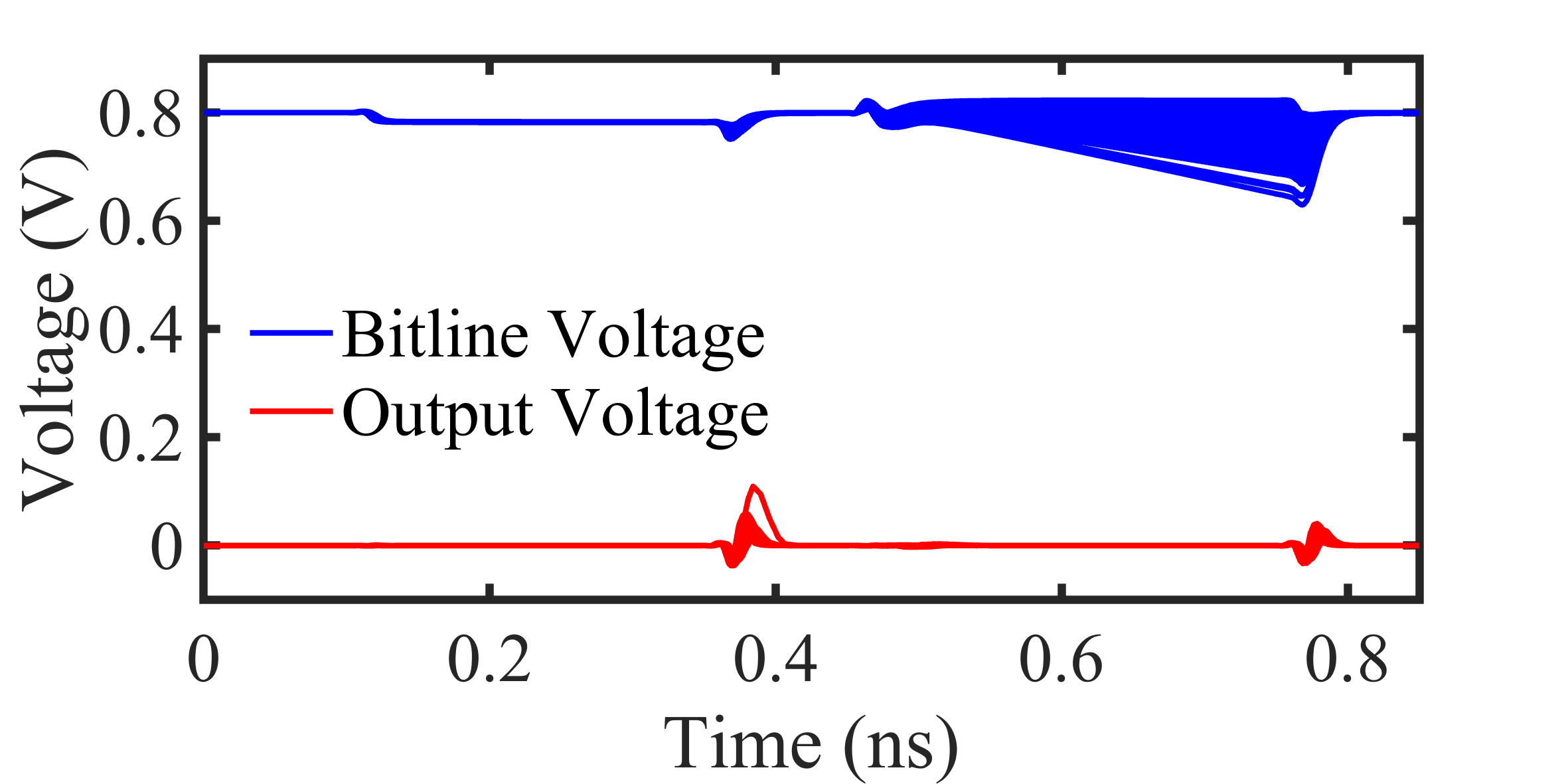}}
\subfloat[case-01]{\includegraphics[width=0.5\linewidth]{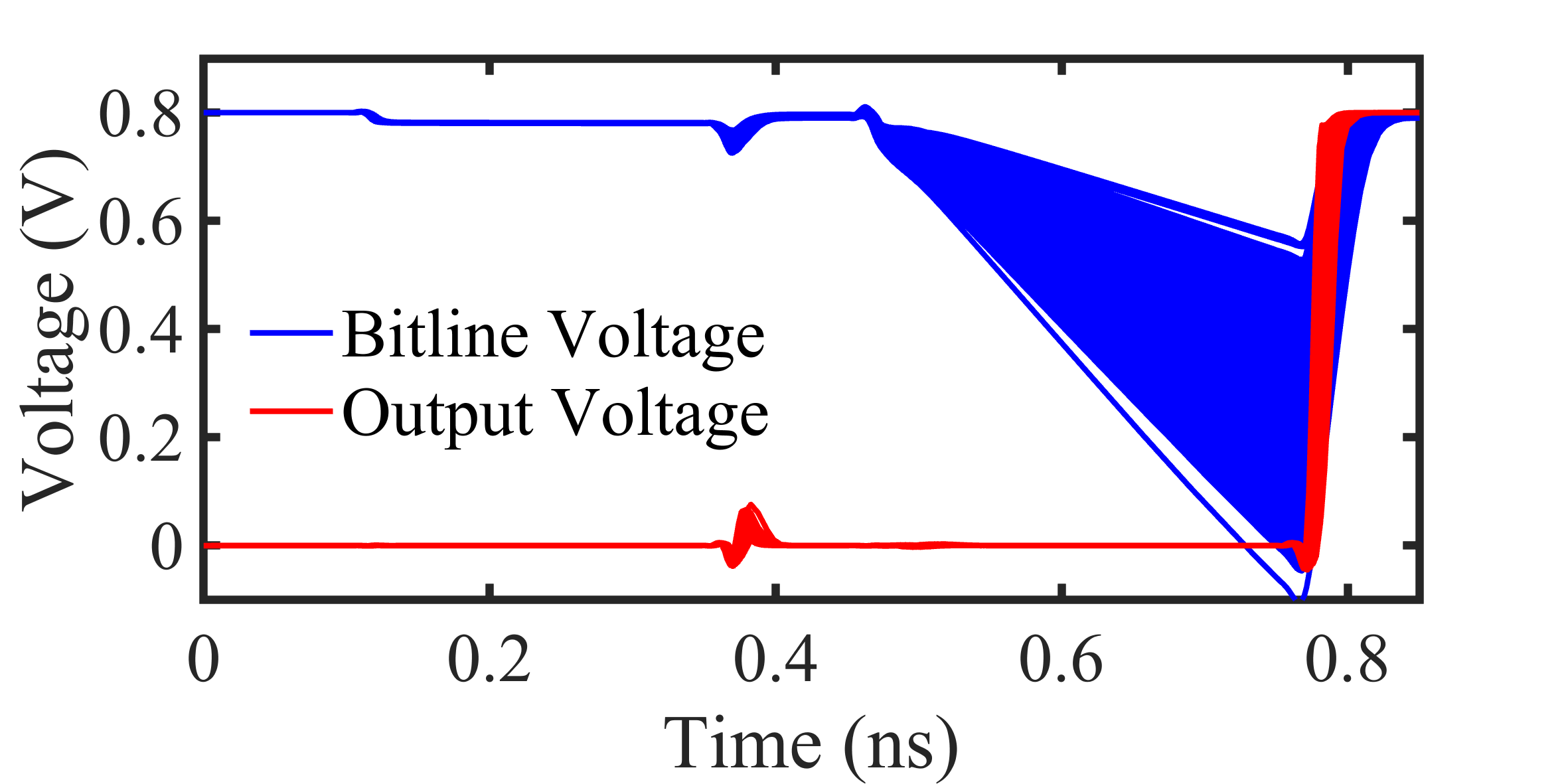}}
\centering
\newline
\subfloat[case-10]{\includegraphics[width=0.5\linewidth]{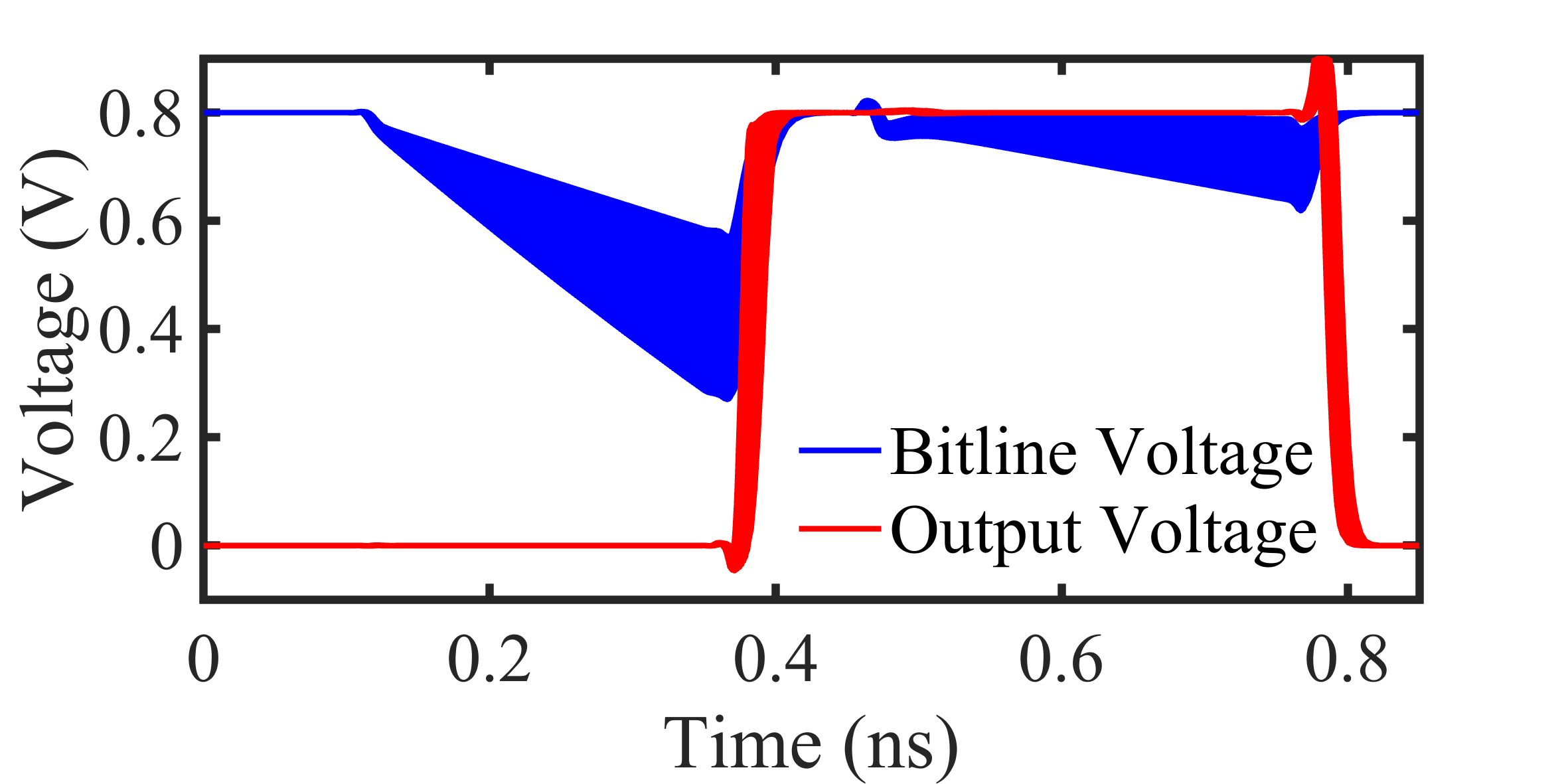}}
\subfloat[case-11]{\includegraphics[width=0.5\linewidth]{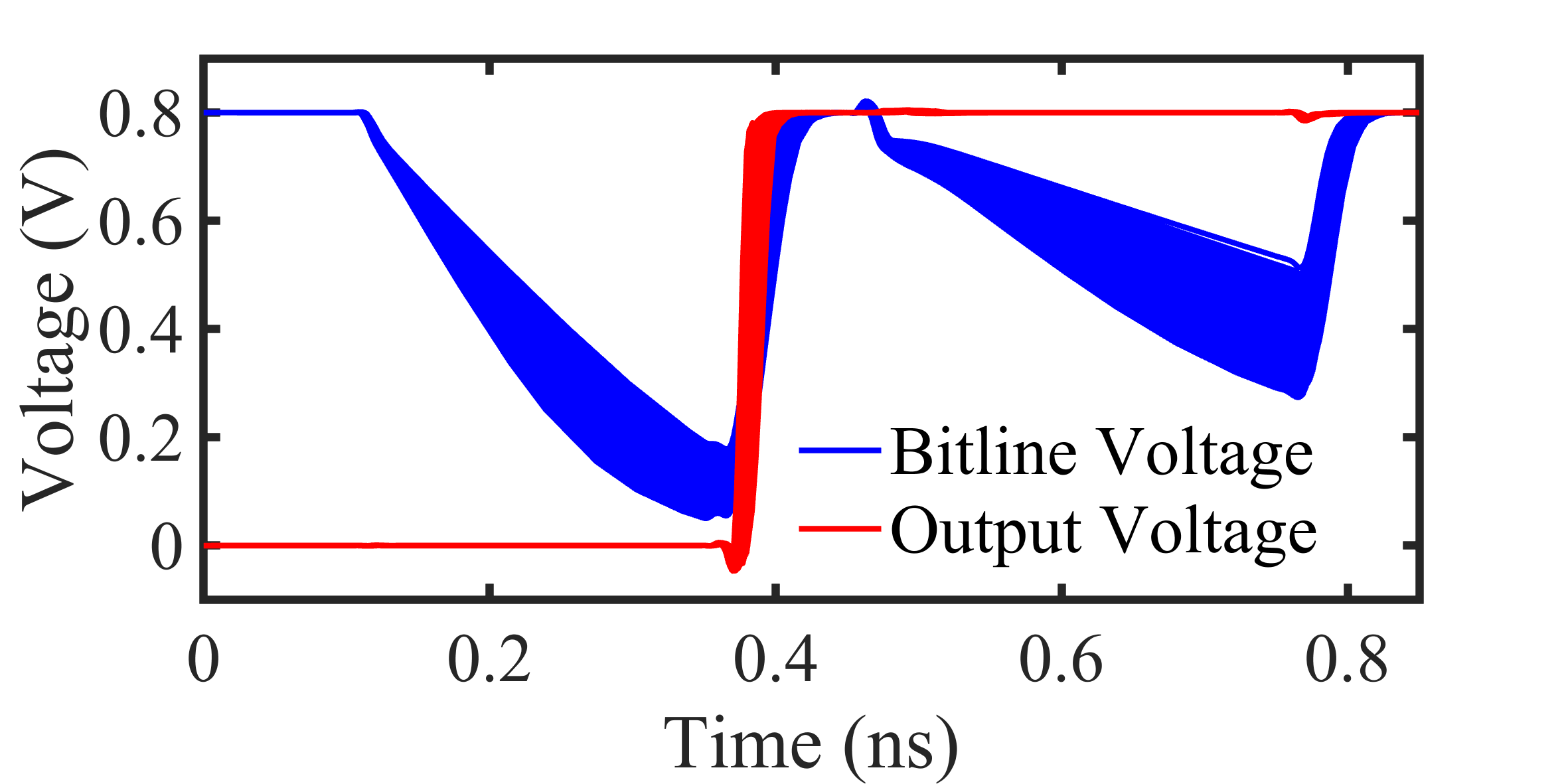}}
\caption{Monte-Carlo simulation results at TT corner for dual-context mode.}
\label{mc_dual_mode}
\end{figure*}

\subsection{Dual-Context Mode}

Figure \ref{mc_dual_mode} shows the Monte-Carlo simulation results for the dual-context mode where both SRAM and ROM data are being read simultaneously, preserving the SRAM data. From the figure, it can be illustrated that the 2-bit data can be sensed robustly for all combinations using one sense amplifier only. In case-00, the output of the sense amplifier remains at the ground constantly. For case-01, the output remains at the ground in the first phase due to the SRAM stored data of `0'; however, it switches to \si{V_{DD}} due to ROM stored data of `1' in the second phase. Accordingly, in case-10, the output first becomes \si{V_{DD}} and goes to the ground in the following phase. Finally, for case-11, the output remains at the \si{V_{DD}} in both phases according to the SRAM and ROM stored data. 

\subsection{Performance Comparison}

Table \ref{tab_ppa} shows the performance metrics of the proposed bit-cell normalized to the standard 8T SRAM bit-cell in both CS and DC modes. The ROM-only mode exhibits higher delay due to gate underdrive (less than \si{V_{DD}}) and high-\si{V_T} read-port transistors. The reliability-friendly RAM-only mode also exhibits higher delay due to the high-\si{V_T} read-port transistors. On the other hand, the delay-friendly RAM-only mode can sense faster than the reliability-friendly RAM-only mode due to the high gate overdrive of the read-port transistors. However, the average read energy in both cases is slightly higher than the standard 8T SRAM read energy. The DC mode shows a higher delay due to the cumulative effect of the SRAM and ROM sensing delay. Since in the DC mode, the timing has to be optimized considering the worst-case scenario (case-10 for the SRAM sensing due to high-\si{V_T} read-port transistors and case-11 for the ROM sensing due to the application of a positive voltage at the source line (SL) to underdrive the gate), delay per bit becomes larger. However, the energy consumption per bit is almost close to the standard 8T SRAM bit-cell. Moreover, the overall leakage of our proposed CS/DC bit-cell is nearly similar to the standard 8T SRAM cell due to the averaging effect from the low-\si{V_T} and high-\si{V_T} read-port transistors. Considering the 6T SRAM bit-cell area and metal pitch for the GF22FDX node, our proposed CS/DC memory exhibits 1.1\si{\times} and 1.3\si{\times} area improvement compared to 6T SRAM and 8T SRAM with separate ROM bank, respectively for iso-memory array size.

\begin{table}[b]
\centering
  \caption{Performance metrics  of the proposed bit-cell normalized to the standard 8T SRAM bit-cell}
  \label{tab_ppa}
  \begin{tabular}{lcccc}
    \hline
    Metrics & ROM-only &  RAM-only &  RAM-only & Dual-Context\\
    &  & Reliability-friendly & Delay-friendly  & \\
    \hline
    Read Delay/bit  & 1.35\si{\times} & 1.79\si{\times} & 1.25\si{\times} & 1.95\si{\times} \\
    Read Energy/bit & 1.06\si{\times} & 1.04\si{\times} & 1.03\si{\times} & 1.08\si{\times} \\
    \hline
    Leakage Power &  \multicolumn{4}{c}{0.997\si{\times}}  \\
    \hline
\end{tabular}
\end{table}

\section{Conclusion}
In this brief, we have presented a context-switching and dual-context ROM augmented 8T SRAM bit-cell. The proposed bit-cell can simultaneously store independent ROM and SRAM data, ensuring the 1.3\si{\times} storage density improvement in the same memory footprint compared to separate RAM and ROM architecture. The memory array can operate as RAM or ROM array (context-switching mode) or RAM and ROM array simultaneously (dual-context mode). The robustness of the functionality of the proposed CS/DC bit-cell has been verified through extensive Monte-Carlo simulations on the GF22FDX technology node. Finally, we believe the proposed bit-cell can be a good candidate for supporting a wide class of emerging applications that need multi-context operations.

\section{Acknowledgments}
The Center for Undergraduate Research at Viterbi (CURVE) partly supported the work. We acknowledge GlobalFoundries's support for 22nm technology.

\bibliographystyle{unsrtnat}
\bibliography{references}

\end{document}